\documentclass[runningheads]{llncs}
\usepackage[T1]{fontenc}
\usepackage{xspace}
\usepackage{amssymb,amsfonts}
\usepackage{graphicx}
\usepackage{textcomp}
\usepackage{xcolor}
\usepackage{soul}
\usepackage{url}
\usepackage{todonotes}
\usepackage{paralist}
\usepackage[inline]{enumitem}
\usepackage{colortbl} 
\usepackage{hyperref}
\usepackage{cleveref}
\usepackage{listings} 
\usepackage{pifont}
\usepackage{multirow}
\usepackage{booktabs}
\usepackage{adjustbox}

\usepackage{epstopdf}
\usepackage{flushend}
\usepackage{tcolorbox}
\usepackage{orcidlink}

\def\BibTeX{{\rm B\kern-.05em{\sc i\kern-.025em b}\kern-.08em
    T\kern-.1667em\lower.7ex\hbox{E}\kern-.125emX}}

\crefformat{section}{§#2#1#3}
\Crefformat{section}{§#2#1#3}

\definecolor{dollarbill}{rgb}{0.52, 0.73, 0.4}
\definecolor{beaublue}{rgb}{0.74, 0.83, 0.9}
\definecolor{ao(english)}{rgb}{0.0, 0.5, 0.0}
\definecolor{amaranth}{rgb}{0.9, 0.17, 0.31}
\definecolor{cerublue}{rgb}{0.16, 0.32, 0.75}
\definecolor{frenchblue}{rgb}{0.0, 0.45, 0.73}
\definecolor{iceberg}{rgb}{0.44, 0.65, 0.82}
\definecolor{blue-violet}{rgb}{0.24, 0.17, 0.99}
\definecolor{darkorchid}{rgb}{0.6, 0.2, 0.8}
\definecolor{my-violet}{rgb}{0.80,0.79,1.00}
\definecolor{ao}{rgb}{0.0, 0.0, 1.0}
\definecolor{atomictangerine}{rgb}{1.0, 0.6, 0.4}
\definecolor{alizarin}{rgb}{0.82, 0.1, 0.26}
\definecolor{americanrose}{rgb}{1.0, 0.01, 0.24}
\definecolor{amber}{rgb}{1.0, 0.75, 0.0}
\definecolor{amber(sae/ece)}{rgb}{1.0, 0.49, 0.0}

\newtcolorbox{quotebox}{colback=dollarbill!50,boxrule=0.4pt,colframe=black,fonttitle=\bfseries,top=2pt,bottom=2pt, before skip=2pt, after skip=2pt, left=2pt,
right=2pt,
top=2pt,
bottom=2pt}

\newcommand{\toolName}{\texttt{GA4GC}}
\newcommand{\repo}{https://github.com/gjz78910/GA4GC}

\title{GA4GC: Greener Agent for Greener Code via Multi-Objective Configuration Optimization}

\titlerunning{GA4GC: Greener Agent for Greener Code}

\author{Jingzhi Gong\inst{1,2}\orcidlink{0000-0003-4551-0701} \and
Yixin Bian\inst{3}\orcidlink{0000-0001-8569-7107} \and
Luis~de~la~Cal\inst{4}\orcidlink{0000-0002-1798-8743} \and
Giovanni Pinna\inst{5}\orcidlink{0000-0003-1362-3322} \and
Anisha Uteem\inst{6}\orcidlink{0009-0000-8036-2367} \and
David Williams\inst{7}\orcidlink{0009-0004-9828-2639} \and
Mar~Zamorano\inst{7}\orcidlink{0000-0002-8872-4876} \and
Karine~Even-Mendoza\inst{6}\orcidlink{0000-0002-3099-1189} \and
W.B.~Langdon\inst{7}\orcidlink{0000-0002-6388-4160} \and
Hector~Menendez\inst{6}\orcidlink{0000-0002-6314-3725} \and
Federica~Sarro\inst{7}\orcidlink{0000-0002-9146-442X}}
\authorrunning{{Jingzhi Gong} {\em et al.}}

\institute{University of Leeds
\email{j.gong@leeds.ac.uk}\and
TurinTech AI \email{jingzhi@turintech.ai}\and
Harbin Normal University
\email{bianyixin@hrbnu.edu.cn}\and
Universidad Politécnica de Madrid
\email{l.delacal@upm.es}\and
University of Trieste
\email{giovanni.pinna@phd.units.it}\and
King's College London
\email{\{anisha.uteem, karine.even\_mendoza, hector.menendez\}@kcl.ac.uk}\and
University College London
\email{\{ucabdjj, maria.lopez.20, w.langdon, f.sarro\}@ucl.ac.uk}
}

\authorrunning{J. Gong et al.}

\begin{document}

\maketitle
\vspace{-0.3cm}
\begin{abstract}
Coding agents powered by LLMs face critical sustainability and scalability challenges in industrial deployment, with single runs consuming over 100k tokens and incurring environmental costs that may exceed optimization benefits. This paper introduces \toolName{}, the first framework to systematically optimize coding agent runtime (greener agent) and code performance (greener code) trade-offs by discovering Pareto-optimal agent hyperparameters and prompt templates. Evaluation on the SWE-Perf benchmark demonstrates up to 135× hypervolume improvement, reducing agent runtime by 37.7\% while improving correctness. Our findings establish temperature as the most critical hyperparameter, and provide actionable strategies to balance agent sustainability with code optimization effectiveness in industrial deployment.
\end{abstract}

\section{Introduction} \label{sec:intro}
Code performance optimization is fundamental to software development, directly impacting system scalability, resource consumption, and user experience~\cite{shypula2023learning}. 
While LLMs show promise in automating this process~\cite{gong2025tuning}, current approaches focus on simple benchmarks like HumanEval~\cite{chen-2021-humaneval} that do not capture real-world software engineering complexity~\cite{gong2025language}. 

To address this limitation, researchers and practitioners have increasingly turned to agentic workflows---sophisticated, multi-step processes where LLMs operate as autonomous agents capable of iterative reasoning, tool use, and complex decision-making~\cite{ashiga2025industrial}. 
These approaches are promising at evaluating realistic SE benchmarks such as 
SWE-Perf~\cite{he2025swe}, which provides code optimization tasks reflecting the complexity that agents face in industry.

However, unlike single-shot LLMs, coding agents operate through iterative reasoning processes that require multiple LLM calls, each consuming significant computational resources~\cite{belcak2025small}. 
While these agents can successfully solve complex real-world coding tasks, a single agent run on real-world SE problems can consume over 100,000 tokens~\cite{anthropic_swe_bench_2025}. Moreover, careful tuning, the energy consumed by an optimization agent can require hundreds of thousands of code executions to reach energetic ``break-even'', making optimization a net energy loss~\cite{coignion2025faster}. As organizations scale deployments, this creates prohibitive costs and threatens environmental sustainability~\cite{iea_datacentres_2020}, directly conflicting with Green Software Engineering principles~\cite{kern2013green} and Net Zero targets~\footnote{\url{https://www.un.org/en/climatechange/net-zero-coalition}}.


\begin{figure} [t!]
    \centering
\includegraphics[width=\linewidth]{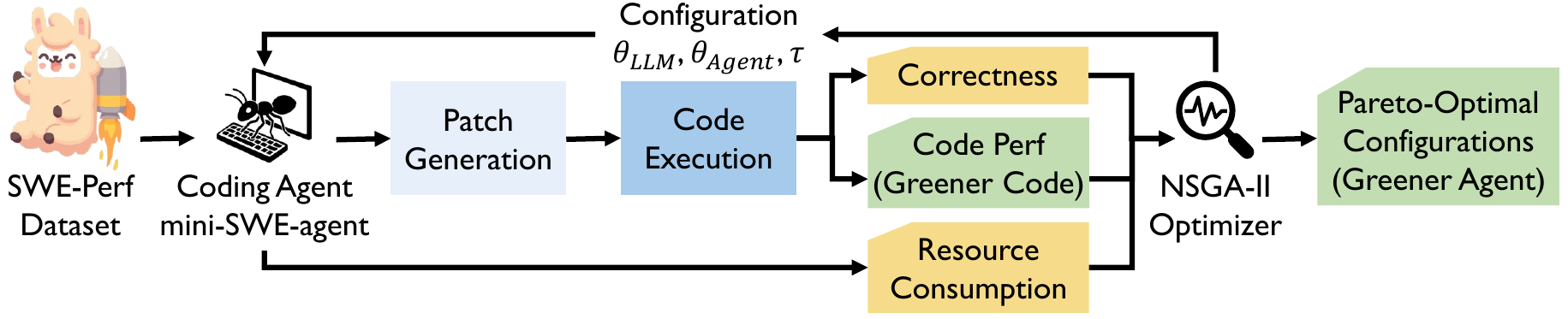}
    \vspace{-0.6cm}
    \caption{\toolName{} workflow of multi-objective configuration optimization.}
    \label{fig:overview}
\vspace{-0.3cm}
\end{figure}

This paper addresses these challenges by proposing \toolName{} (Greener Agent for Greener Code), which optimizes the trade-off between resource consumption of the coding agent and performance of the generated code. Our key insight is that the vast configuration space of coding agents---including prompt 
templates, LLM, and agent-specific hyperparameters---is too complex for manual exploration. Thus, \toolName{} employs NSGA-II 
multi-objective optimization (MOGA) to discover Pareto-optimal agent configurations. \textbf{Our contributions are}:
\begin{itemize}
    \item \toolName{}, a MOGA framework that discovers Pareto-optimal coding agent configurations that are up to 37.7\% faster (943.1s vs 1513.3s) while improving correctness, and with up to 135× improved hypervolume over the default.
    \item Hyperparameter influence analysis revealing that temperature is critical for code performance (0.392), timeout constraints improve agent efficiency, and top\_p/cost\_limit create performance-runtime trade-offs.
    \item Actionable suggestions for green SBSE practitioners across three scenarios: runtime-focused (Config\#4, 37.7\% reduction), performance-focused (Config\#15, 10.67\% improvement), and balanced (Config\#5, comprehensive gains), with \toolName{} enabling context-specific optimization.
\end{itemize}

\textbf{Related Work.}
Recent green GenAI research has applied reinforcement learning for energy-efficient code generation~\cite{ilager2025green}, compared energy efficiency of LLM- versus human-written code~\cite{apsan2026energy}, and optimized GenAI hyperparameters for domain modeling~\cite{bulhakov2025investigating} and text-to-image generation~\cite{gong2024greenstableyolo}.
These approaches, however, focus on single-shot generative tasks. By contrast, we address the challenges of complex, multi-turn agentic workflows, mitigating the substantial computational costs of deploying coding agents in real-world software engineering.

\section{Methodology and Experimental Setup}

\noindent \textbf{MOGA Optimization.}
\Cref{fig:overview} illustrates \toolName{}'s workflow, where we employ NSGA-II to explore the {agent configuration space defined by $\mathcal{C} = (\theta_{LLM}, \theta_{agent}, \tau)$}, where $\theta_{LLM}$ represents LLM-specific hyperparameters, $\theta_{agent}$ represents agent-specific operational constraints, and $\tau$ represents the prompt template variant \footnote{Details on the prompts we used can be found in our \href{\repo}{replication package}.}. \Cref{tab:hyperparameters} details the configuration search space.

We define three fitness functions: \textbf{$f_1(\mathcal{C})=$correctness (passes all test cases), $f_2(\mathcal{C})=$performance gain (code speedup), and $f_3(\mathcal{C})=$agent runtime (to minimize)}. For each candidate configuration, the agent receives a code optimization task and generates patches through iterative reasoning, during which we measure $f_3$. Generated patches are executed in isolated Docker environments to measure $f_1$ and $f_2$, and the output is a Pareto front of non-dominated configurations.



\vspace{0.3cm}
\noindent \textbf{Research Questions.}
We address three research questions (RQs):
\vspace{-0.3cm}
\begin{description}
    \item[\ding{228} RQ1.] To what extent can \toolName{} improve the resource consumption and performance trade-offs of coding agents compared to default configurations?
    \item[\ding{228} RQ2.] How do different hyperparameters influence agent resource consumption and task performance in the optimization process?
    \item[\ding{228} RQ3.] What actionable strategies can be derived from the Pareto-optimal configurations for sustainable coding agent deployment?
\end{description}



\begin{table}[t!]
\centering
\caption{Configuration search space (decimal range $=$ any value within the range; integer range $=$ only integer values; set $=$ only specified values).}
\vspace{-0.2cm}
\begin{adjustbox}{width=\columnwidth,center}
\begin{tabular}{lllcl}
\toprule
\textbf{Category} & \textbf{Hyperparameter} & \textbf{Abbr.} & \textbf{Range/Values} & \textbf{Description} \\
\midrule
\multirow{3}{*}{LLM} 
 & Temperature    & Temp  & [0.0, 1.0]       & Controls randomness in token selection \\
 & Top\_p         & TopP  & [0.1, 1.0]       & Limits sampled token vocabulary size \\
 & Max\_tokens    & Token & [512, 4096]      & Constrains maximum response length \\
\hline
\multirow{4}{*}{Agent} 
 & Step\_limit    & Step  & [10, 40]         & Limits number of LLM calls \\
 & Cost\_limit (\$)& Cost & [3.0, 10.0]      & Constrains total cost of LLM usage \\
 & Env\_timeout (s)& ETi  & [40, 60]         & Timeout for environment operations \\
 & LLM\_timeout (s)& LTi  & [40, 60]         & Timeout for individual LLM calls \\
\hline
Prompt & Template Variant & Pr & \{1,2,3\} & Different template configurations \\
\bottomrule
\end{tabular}
\end{adjustbox}
\label{tab:hyperparameters}
\vspace{-0.3cm}
\end{table}




\noindent \textbf{Experimental Setup.}
We use mini-SWE-agent~\cite{miniSWEagent2024} with Gemini 2.5 Pro as the base LLM. 
The evaluation employs SWE-Perf~\cite{he2025swe}, a benchmark for code optimization tasks in real-world repositories where the goal is to improve code runtime while maintaining functionality. Given the extensive evaluation time required for each candidate configuration, we focus on the astropy project, using 9 instances for NSGA-II optimization and 3 instances for validation. 

NSGA-II explores the configuration space over 5 generations with population size 5, evaluating 25 total configurations (25-35 hours and \$50-100 LLM API costs per run). We use pymoo's default NSGA-II setup: binary tournament selection, simulated binary crossover with probability 0.9, and polynomial mutation with probability $1/\textit{n\_vars}$. Each configuration is evaluated by running the agent on all 9 training instances, measuring the three objectives ($f_1$, $f_2$, $f_3$). After optimization, we extract the Pareto-optimal configurations and validate them on 3 held-out instances to assess generalization. 

All experiments are conducted on an isolated Google Cloud Platform server with 4 CPUs, 16GB RAM, running Ubuntu 25.04. Performance gains for each SWE-Perf instance are measured 20 times, and statistical significance is evaluated using the Mann-Whitney U test with $p < 0.1$. 
\section{Results and Analysis} \label{sec:results}

\begin{table}[t!]
\centering
\rowcolors{2}{gray!10}{white}
\caption{Comparison between default and \toolName{}-optimized configurations. RT=runtime, HV=hypervolume, VHV=validation hypervolume. See \Cref{tab:hyperparameters} for other definitions. \setlength{\fboxsep}{1.5pt}\colorbox{dollarbill!30}{Green cells} indicate improvements over default. }
\vspace{-0.2cm}
\begin{adjustbox}{width=\columnwidth,center}
\begin{tabular}{l|cccccccc|ccc|cc}
\toprule
\textbf{Config} & \textbf{Temp} & \textbf{TopP} & \textbf{Token} & \textbf{Step} & \textbf{Cost} & \textbf{ETi} & \textbf{LTi} & \textbf{Pr} & \textbf{Corr} & \textbf{Perf (\%)} & \textbf{RT (s)} & \textbf{HV (\%)} & \textbf{VHV (\%)} \\
\midrule
Default & 0.0 & 1.0 & 4096 & 240 & 3.0 & 60 & 60 & - & 2/9 & 0.00 & 1513.3 & 0.52 & 1.1 \\
\#4 & 0.085 & 0.135 & 1120 & 36 & 9.26 & 41 & 57 & 2 & \cellcolor{dollarbill!50}4/9 & 0.00 & \cellcolor{dollarbill!50}943.1 & \cellcolor{dollarbill!50}5.82 & \cellcolor{dollarbill!50}4.1 \\
\#5 & 0.692 & 0.384 & 2972 & 38 & 6.73 & 40 & 56 & 3 & \cellcolor{dollarbill!50}8/9 & \cellcolor{dollarbill!50}6.43 & \cellcolor{dollarbill!50}984.8 & \cellcolor{dollarbill!50}70.28 & \cellcolor{dollarbill!50}14.9 \\
\#9 & 0.725 & 0.412 & 2972 & 22 & 6.73 & 43 & 41 & 3 & \cellcolor{dollarbill!50}7/9 & 0.00 & \cellcolor{dollarbill!50}958.1 & \cellcolor{dollarbill!50}9.25 & \cellcolor{dollarbill!50}21.6 \\
\#15 & 0.657 & 0.384 & 2972 & 38 & 6.73 & 40 & 56 & 2 & \cellcolor{dollarbill!50}7/9 & \cellcolor{dollarbill!50}10.67 & \cellcolor{dollarbill!50}1400.1 & \cellcolor{dollarbill!50}33.42 & \cellcolor{dollarbill!50}2.7 \\
\#16 & 0.085 & 0.131 & 1120 & 36 & 6.91 & 41 & 57 & 2 & 0/9 & 0.00 & \cellcolor{dollarbill!50}853.3 & \cellcolor{dollarbill!50}1.10 & \cellcolor{dollarbill!50}21.6 \\
\bottomrule
\end{tabular}
\end{adjustbox}
\label{tab:rq1_performance}
\vspace{-0.6cm}
\end{table}

\noindent \textbf{RQ1 Results.}
\Cref{tab:rq1_performance} shows the results of RQ1, where NSGA-II identifies five Pareto-optimal configurations: Config\#4 achieves 37.7\% runtime reduction (943.1s vs 1513.3s) while doubling correctness, Config\#15 achieves 10.67\% code performance improvement with similar runtime overhead, and Config\#5 delivers 4× better correctness (8.0 vs 2.0) while simultaneously improving performance by 6.43\%. Notably, \textbf{four out of five configurations dominate in multiple objectives} \footnote{Pareto front visualizations and baseline comparison are available in our \href{\repo}{GitHub}.}, addressing both greener agent and greener code requirements.

We computed the hypervolume indicator using \texttt{pymoo} with objectives normalized to [0,1] and reference point [-0.1, -0.1, -0.1] (runtime inverted). \textbf{Each optimized configuration substantially outperforms the default}: Config\#5 achieves 135× higher hypervolume (70.28\% vs 0.52\%), Config\#15 achieves 64× improvement (33.42\% vs 0.52\%), and even the lowest-performing Config\#16 achieves 2× improvement (1.10\% vs 0.52\%). Validation on three held-out instances confirms generalization, with all optimized configurations maintaining superior hypervolume.

\begin{quotebox}
   \noindent
\textbf{RQ1:} \toolName{} achieves 135× higher hypervolume, 37.7\% faster runtime while improving correctness, and 4/5 Pareto front configurations dominating the default while all maintaining superior hypervolume on unseen tasks.
\end{quotebox}

\vspace{0.3cm}
\noindent \textbf{RQ2 Results.}
\Cref{tab:rq2_hyperparams} shows the hyperparameter influence analysis. We train a Random Forest for each objective using all 25 evaluated configurations to measure influence magnitudes \cite{gong2024greenstableyolo}. Among others, \textbf{temperature emerges as the most critical hyperparameter}, with high-performing Config\#5 and \#15 using moderate temperatures (0.66-0.69) while low-temperature Config\#4 and \#16 achieve faster runtime but no performance gain, indicating its role in balancing exploration versus exploitation during token generation. 

Top\_p shows correctness influence (0.199) with successful configurations using mid-range values (0.38-0.41), indicating that balanced vocabulary sampling avoids both overly restrictive and chaotic token selection. Cost\_limit exhibits influence across correctness (0.199) and runtime (0.128), with Pareto-optimal configurations using higher budgets (\$6.73-\$9.26 vs \$3.0 default) to enable more thorough exploration without timeout constraints. Prompt template variants show moderate performance influence (0.130), with templates 2 and 3 dominating the Pareto front, suggesting that task-specific prompt engineering significantly impacts optimization effectiveness. 
\begin{quotebox}
   \noindent
\textbf{RQ2:} Temperature shows highest overall influence, LLM hyperparameters primarily impact task effectiveness while agent constraints affect resource consumption, confirming the need for MOGA in green coding agent deployment.
\end{quotebox}

\begin{table}[t!]
    \centering
    \caption{Random Forest feature importance for hyperparameters on optimization objectives. Colors indicate importance: \setlength{\fboxsep}{1.5pt}\colorbox{amber!35}{Low (0.0-0.1)}, \setlength{\fboxsep}{1.5pt}\colorbox{amber(sae/ece)!45}{Medium (0.1-0.2)}, \setlength{\fboxsep}{1.5pt}\colorbox{red!55}{High (>0.2)}.}
    \begin{adjustbox}{width=\columnwidth,center}
    \begin{tabular}{llccc}
    \toprule
    \textbf{Category} & \textbf{Hyperparameter} & \textbf{Correctness Impact} & \textbf{Performance Impact} & \textbf{Runtime Impact} \\
    \midrule
    \multirow{3}{*}{LLM} & Temperature & \cellcolor{amber(sae/ece)!50}0.152 & \cellcolor{red!60}0.392 & \cellcolor{amber(sae/ece)!50}0.199 \\
    & Top\_p & \cellcolor{amber(sae/ece)!50}0.199 & \cellcolor{amber!40}0.051 & \cellcolor{amber!40}0.097 \\
    & Max\_tokens & \cellcolor{amber!40}0.057 & \cellcolor{amber!40}0.090 & \cellcolor{amber!40}0.089 \\
    \hline
    \multirow{4}{*}{Agent} & Step\_limit & \cellcolor{amber(sae/ece)!50}0.140 & \cellcolor{amber(sae/ece)!50}0.119 & \cellcolor{amber!40}0.049 \\
    & Cost\_limit & \cellcolor{amber(sae/ece)!50}0.199 & \cellcolor{amber!40}0.076 & \cellcolor{amber(sae/ece)!50}0.128 \\
    & Env\_timeout & \cellcolor{amber!40}0.060 & \cellcolor{amber!40}0.034 & \cellcolor{red!60}0.298 \\
    & LLM\_timeout & \cellcolor{amber(sae/ece)!50}0.120 & \cellcolor{amber(sae/ece)!50}0.109 & \cellcolor{amber(sae/ece)!50}0.102 \\
    \hline
    Prompt & Template Variant & \cellcolor{amber!40}0.072 & \cellcolor{amber(sae/ece)!50}0.130 & \cellcolor{amber!40}0.038 \\
    \bottomrule
    \end{tabular}
    \end{adjustbox}
    \label{tab:rq2_hyperparams}
    \vspace{-0.3cm}
    \end{table}

\vspace{0.3cm}
\noindent \textbf{RQ3 Results.}
Based on the hyperparameter influence analysis, we derive actionable strategies for green SBSE practitioners across different optimization scenarios:
(1) \textbf{For runtime-critical scenarios:} Use low temperature (0.0-0.1) with restrictive top\_p (0.13-0.14) to minimize exploration overhead, combined with moderate max\_tokens (1120-2000) and step limits (20-36). 
(2) \textbf{For performance-critical scenarios:} Use moderate temperature (0.65-0.70) with balanced top\_p (0.38-0.41) to enable creative optimization strategies, combined with higher cost budgets (\$6.5-\$9.5) and prompt templates optimized for performance tasks.
(3) \textbf{For most accurate optimization:} For practitioners with specific requirements, we recommend applying \toolName{} to discover context-specific Pareto-optimal configurations tailored to their deployment priorities. 

\begin{quotebox}
   \noindent
\textbf{RQ3:} We provide scenario-specific actionable suggestions for green SBSE practitioners. For more accurate optimization, practitioners can apply \toolName{} to discover context-specific Pareto-optimal configurations.
\end{quotebox}

\vspace{0.3cm}
\noindent \textbf{Threats to Validity.}
 Our evaluation focuses on the astropy project (12 instances) from SWE-Perf and are specific to mini-SWE-agent with Gemini 2.5 Pro due to computational constraints, which may limit generalizability. The limited search budget may prevent full Pareto front convergence. The stochastic nature of NSGA-II and LLM inference (with non-zero temperature) means results may vary across runs. All limitations reveal opportunities for future studies.  
\section{Conclusion}

This paper introduced \toolName{}, a framework to optimize coding agent resource-performance trade-offs via multi-objective optimization. On SWE-Perf, it achieves 135× hypervolume improvement and 37.7\% runtime reduction while improving correctness. Our analysis also reveals insights and actionable guidelines to address both green computing concerns and industrial deployment requirements. 

\vspace{0.3cm}
\noindent\textbf{Availability.}
Code and results are available at \href{\repo}{GitHub} \& \href{https://doi.org/10.5281/zenodo.17177692}{\raisebox{-0.4ex}{\includegraphics[height=1.1em]{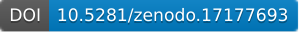}}}.


\bibliographystyle{splncs04}
\bibliography{references}

\end{document}